\documentclass[doublecol]{epl2}
\usepackage{amsmath}
\usepackage{latexsym}
\usepackage{amssymb}
\usepackage{subfigure}
\usepackage{color}

\definecolor{linkcolor}{rgb}{0,0,1} 

\title{Fluctuations, linear response and heat flux of an aging system} 

\author{J. R. Gomez-Solano\inst{1}\footnote{Current address: 2. Physikalisches Institut, Universit\"at Stuttgart, Pfaffenwaldring 57, 70569, Stuttgart, Germany.} \and
  A. Petrosyan\inst{1} \and S. Ciliberto\inst{1}}
\shortauthor{J. R. Gomez-Solano \etal}

\institute{
  \inst{1} Laboratoire de Physique, Ecole Normale Sup\'erieure de Lyon,
CNRS UMR 5672, 46, All\'ee d’Italie, 69364 Lyon CEDEX 07, France\\
}
\pacs{05.40.-a}{Fluctuation phenomena, random processes, noise, and Brownian motion}

\abstract{We measure the fluctuations of the position of a Brownian particle confined by an optical trap in an aging gelatin droplet after a fast quench. Its linear response to an external perturbation is also measured. We compute the spontaneous heat flux from the particle to the bath due to the nonequilibrium formation of the gel. We show that the mean heat flux is quantitatively related to the violation of the equilibrium fluctuation-dissipation theorem as a measure of the broken detailed balance during the aging process.}

\begin{document}

\maketitle

\section{Introduction}
The study of the statistical properties of fluctuations in out-of-equilibrium systems is a topic of great current interest. For instance, it is important at micro- and nano- scales where fluctuations and off-equilibrium conditions are common. 
In particular, the fluctuation theorem  (FT) \cite{gallavotti} and the generalization of the fluctuation-dissipation theorem (FDT) \cite{Seifert_2010,cugliandolo,lippiello,baiesi0,verley} to processes away from thermal
equilibrium, are  important theoretical results relevant to describe energy exchanges. 
A large number of studies has been devoted to non-equilibrium steady states and transient 
states  in contact with a single heat bath. Nowadays,
these systems begin to be rather well understood both theoretically and experimentally 
(see ref.\cite{ciliberto} for a short review). On the other hand, non-stationary states slowly relaxing towards thermal equilibrium are important for applications. For example, glassy systems age in time driven by spontaneous heat fluxes to the environment, after being prepared in a metastable configuration \cite{berthier}. Although several extensions of the FT \cite{crisanti,ritort,zamponi,chetrite} and  generalized fluctuation-dissipation relations (GFDRs) \cite{cugliandolo,lippiello,baiesi0,verley} have been formulated for relaxing systems, the comparison between theory and experiments still lacks a clear interpretation. 

In the present letter we  study experimentally the fluctuations, the linear response and the heat flux of a non-stationary system: an  aging gel during the sol-gel transition. The rheological properties and the fluctuations during this aging dynamics can be 
characterized  by a single relevant degree of freedom of a local probe. Specifically, we focus, within  the context of GFDRs, on the dynamics of a micron-sized particle embedded in the aging gel, which acts as a non-equilibrium bath. The particle is the probe, which measures  the out-of-equilibrium properties of the gel. Our aim is to show in a transparent way the link between the violation of the equilibrium FDT and the total entropy production in an experiment performed in an aging system.

\section{Description of the system}
In the experiment, a silica bead of radius $r=1\,\mu$m is used as a probe inside a thermoreversible gel (gelatin) slowly relaxing
towards its solid-like state (gel), after a very fast quench, from
above to below the gelation temperature $T_{gel}$. Above
$T_{gel}$ an aqueous gelatin solution is in a viscous liquid phase
(sol), whereas below $T_{gel}$ the formation of a network of
cross-linked triple helices leads to an elastic solid-like phase (gel)
\cite{djabourov1}. 
This relaxational out-of-equilibrium regime is probed by measuring the Brownian fluctuations and the linear response function of the bead position. 

Specifically, the gel used in the experiment is 
an aqueous gelatin solution (type-B pig skin) at a concentration of 10~wt~\%, prepared following the usual
protocol \cite{normand,gomez}. For this sample $T_{gel} = 29^{\circ}$C.
The solution fills a transparent cell kept at constant temperature ($T_0=26\pm 0.05^\circ$C $<T_{gel}$)
by a Peltier element. See the sketch of the experimental setup on the left panel of fig.~\ref{fig:quenchmicro}(a). Thus, the solution inside the cell is in the solid-like phase.  A silica bead, of radius
$r=1\,\mu$m, embedded the gelatin solution is placed in the focal position of a
tightly focused laser beam ($\lambda=980$ nm) at a power of 20 mW.
At this power the laser produces on the particle an elastic force
of stiffness $k=2.9 \ \rm{pN}/\mu$m. Because of light absorption,
the temperature of the trapped particle is $T=27.2^\circ$C, which is
still smaller than $T_{gel}$. Hence, the bead is inside the
solid gel in the beam focus at a distance $h=25\,\mu$m from the
cell wall. Starting from this condition, the laser power is increased to 200 mW and the
local temperature around the focus rises to $38^{\circ}$C
$>T_{gel}$ { \cite{peterman}}. As a result the gel melts and a liquid droplet of
radius $a=5\,\mu$m, is formed around the trapped bead inside the
the solid gel bulk, as depicted on the right panel of fig.~\ref{fig:quenchmicro}(a).
After 180~s, the laser power is suddenly decreased again to 20~mW and  the temperature is homogenized by heat diffusion into the gel bulk. Since the thermal diffusivity of water is $\kappa = 1.4 \times 10^{-7}\,\mathrm{m}^2 \, \mathrm{s}^{-1}$,  the heat diffuses in a time $a^2/\kappa \sim 0.2 $~ms and the droplet is very efficiently quenched to the final temperature $T=27.2^{\circ}\mathrm{C}<T_{gel}$.  { This temperature remains constant after the quench due the Peltier element at $T_0=26\pm 0.05^\circ$C. In addition, we check that the trap stiffness $k =2.9 \ \rm{pN}/\mu$m remains constant as well throughout the gelation process.}
At $T$ the liquid inside the droplet completely solidifies in about 1 hour and the
particle, trapped in the center of the drop by the focused beam,
is a probe of this relaxation dynamics. {    The quenches are repeated up to  60  times in order to  compute the proper ensemble
averages of the measured quantities \cite{Solano_thesis}}.

\section{Active microrheology}
\begin{figure}
     \centering
        {  \includegraphics[width=.35\textwidth]{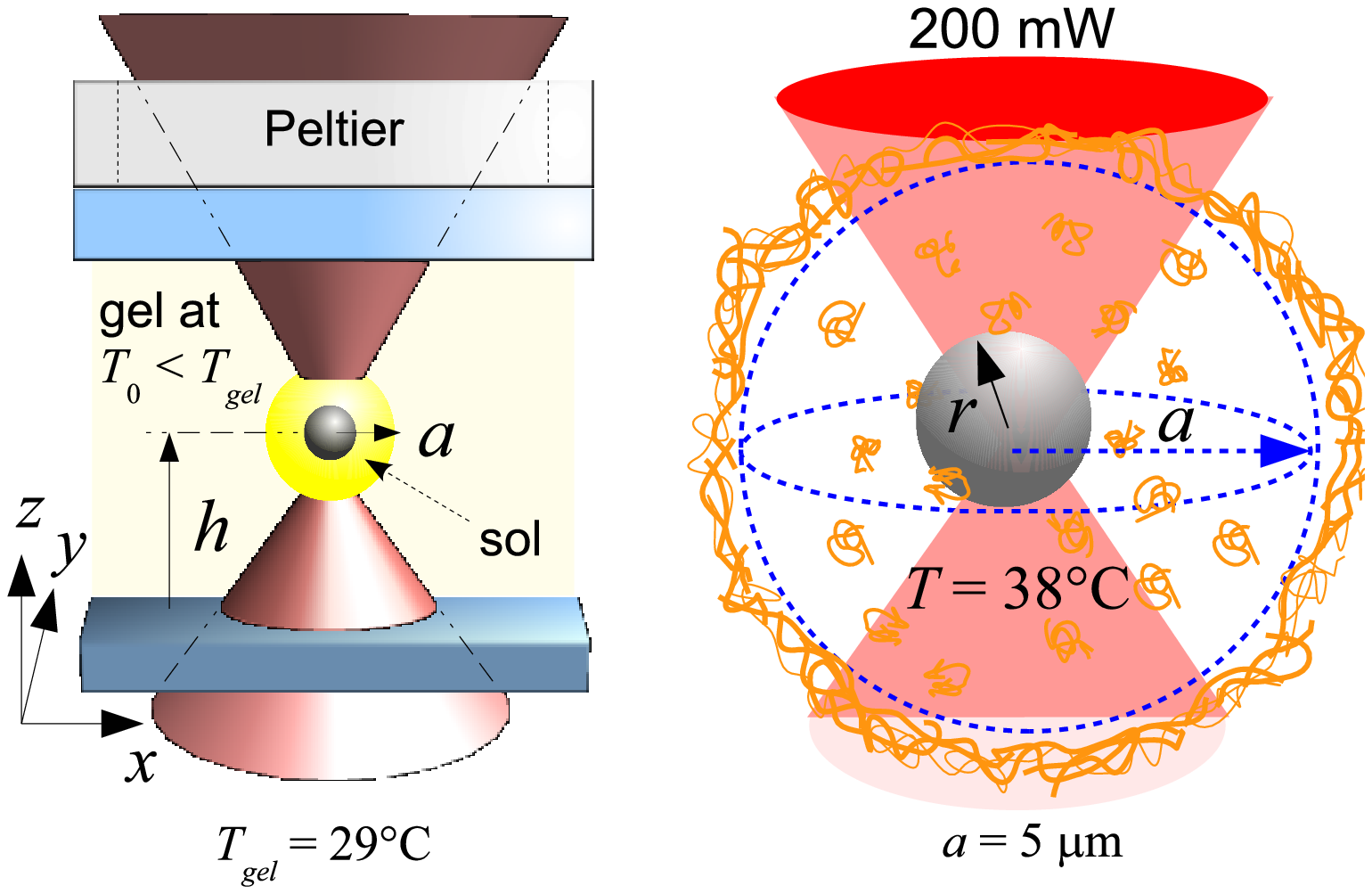}}\\
    \vspace{-0.2in}
         { \includegraphics[width=.35\textwidth]{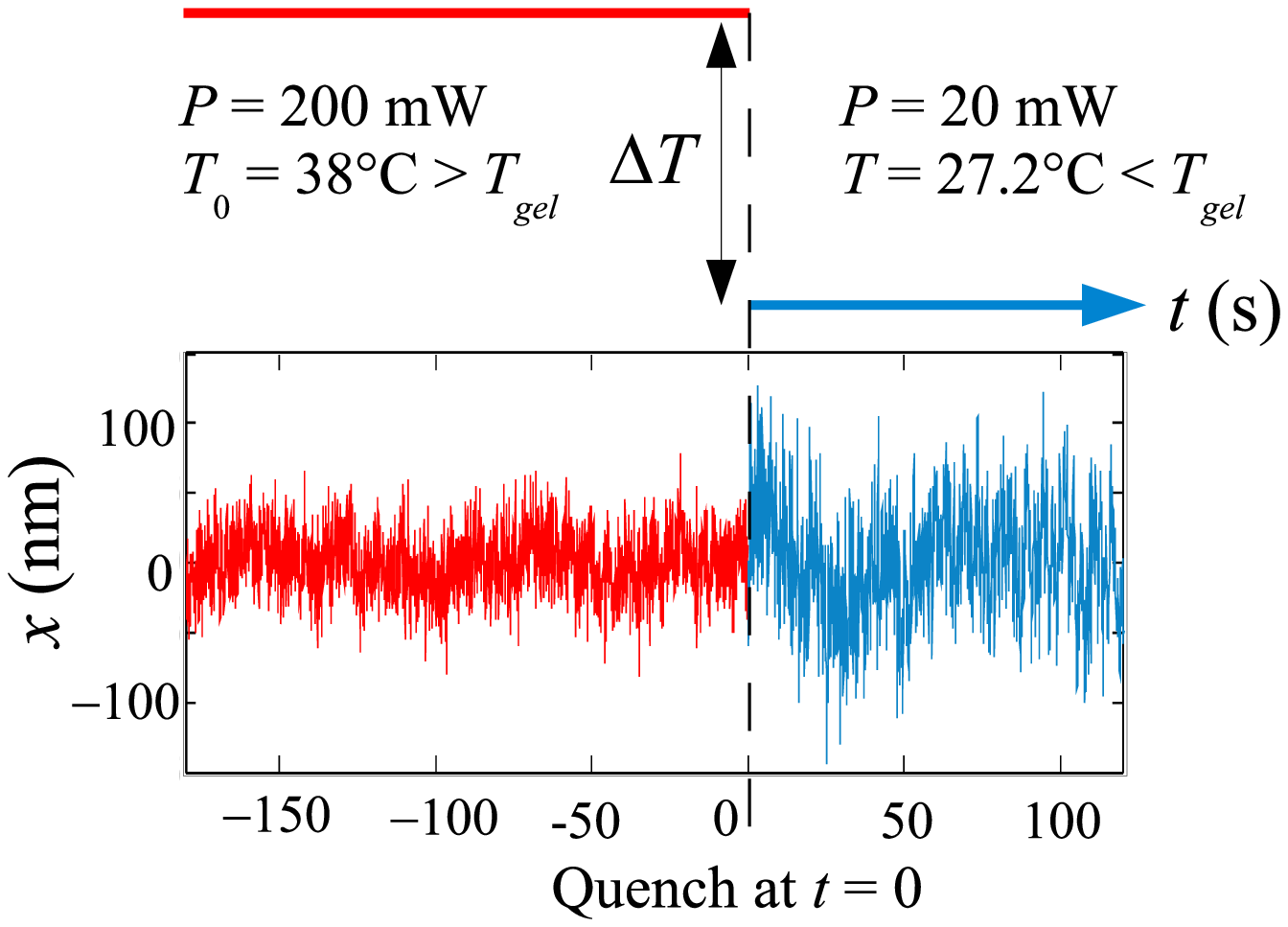}}
     \caption{(a) Sketch of the experimental setup used to melt a micron-sized volume of gel by locally heating the bulk at $T_0=38^{\circ}$C with a focused laser beam. (b) Diagram of the quench of the liquid droplet from $T_0=38^{\circ}$C to $T=27.2^{\circ}$C in less than $1$~ms. Lower panel: position fluctuations of a particle optically trapped inside the liquid before and after the quench.}
     \label{fig:quenchmicro}
\end{figure}
Immediately after the quench we record the time evolution of the
$x$ position [see fig.~\ref{fig:quenchmicro}(b)] of the trapped particle
measured by a position sensitive detector whose output is sampled
at 8 kHz and acquired by a computer. The resolution of the
measurement of $x$ is better than $1$ nm \cite{resolution,jop}. 

Due to the aging process of the gelatin droplet, the time series of $x$ shown in fig.~\ref{fig:quenchmicro}(b) is non-stationary for time $t \ge 0$ after the quench. 
In order to characterize this aging time evolution in the \emph{linear response regime}, we perform active microrheology. For this purpose we apply an { external driving to the trapped bead. This is realized by displacing  the position $x_0$ of the optical trap, using a precise  actuator, which deflects the  trapping laser beam. A sinusoidal modulation of $x_0$ results in a time-dependent force $F = k x_0$ on the bead: $F(f,t) = F_0 \sin(2\pi \ f \ t)$ is  applied at different driving frequencies 0.2~Hz~$\le f \le$~5~Hz and fixed amplitude $F_0=87$~fN. }
{\color{red} By measuring in the time interval [$t,t+\Delta t$] the response $\hat R (f,t)$ of the particle position to $F(f, t)$, we obtain the shear modulus {  $G(f,t)=G'(f,t)-\mathrm{i}G''(f,t)$} of the droplet at different times $t$ after the quench, where $G'$ is the storage and $G''$ the loss modulus \cite{jop} (see also eq.\ref{eq:RG})}.

\begin{figure}
     \centering
     \subfigure[]{
          \includegraphics[width=.252\textwidth]{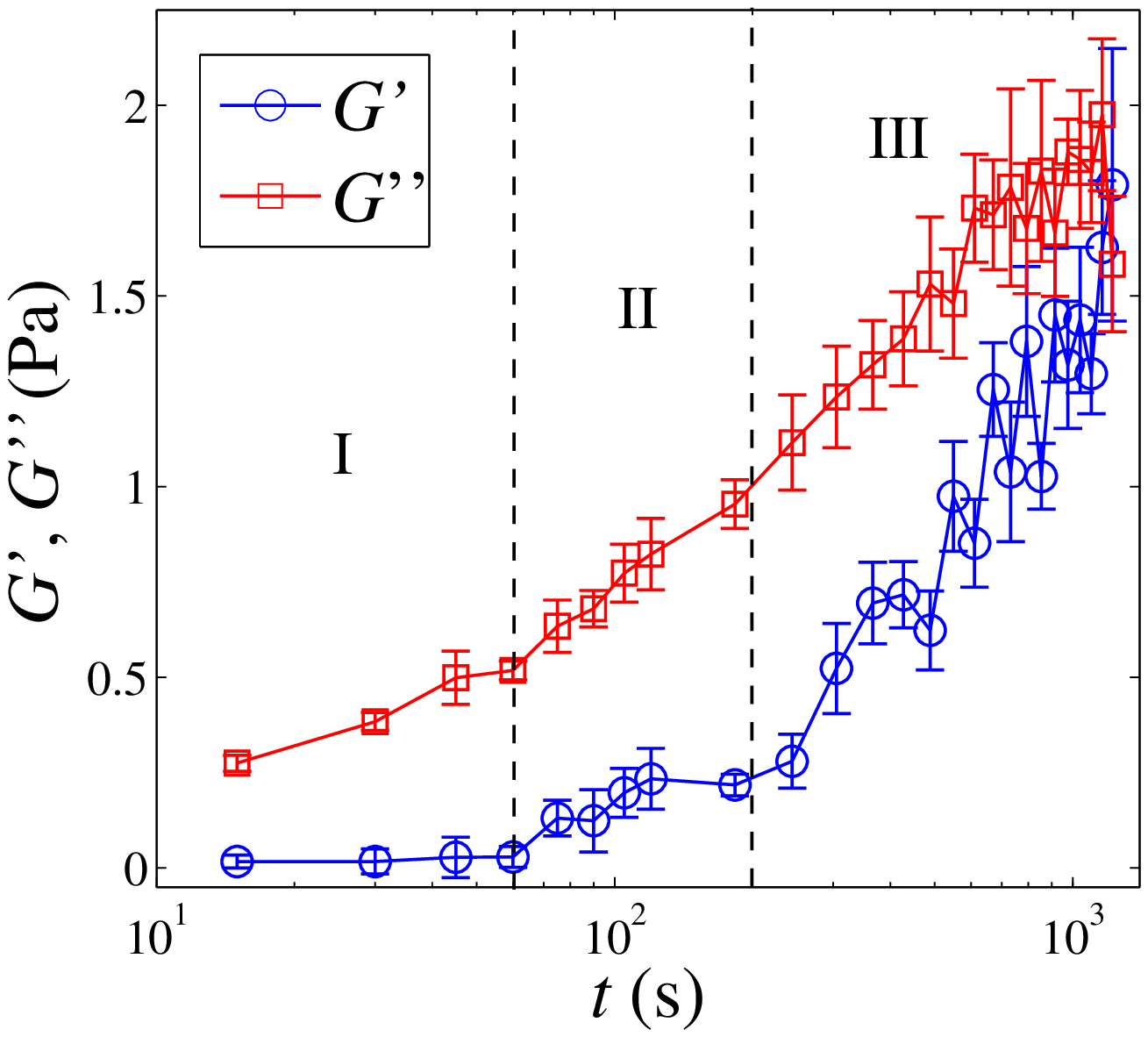}}
     \hspace{-0.275in}
     \subfigure[]{
          \includegraphics[width=.252\textwidth]{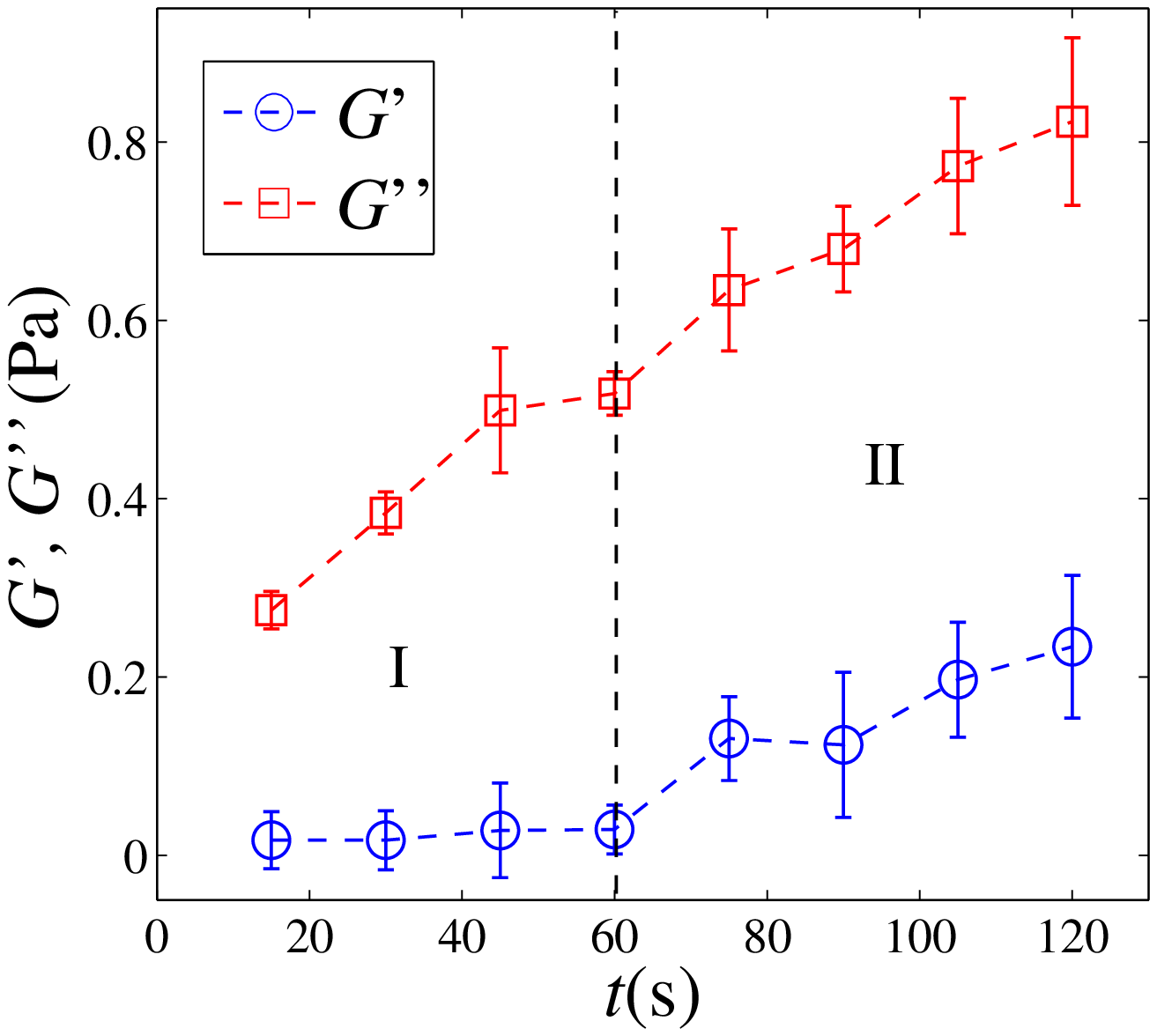}}\\
     \caption{(a)  Time evolution of $G'$ and $G''$ of the gelatin droplet after the quench, measured at $f=5$ Hz. (b) Expanded view during the first 120~s. { The error bars represent the standard deviation computed from 20 independent quenches. }} 
     \label{fig:GGel}
\end{figure}

Because of the relatively fast aging process due to the smallness of the droplet, the spectral analysis involved in the calculation of $G'$ and $G''$ must be carried out over a short time window $[t,t+\Delta t]$ for each aging time $t$. In the following we set $\Delta t=15$~s\footnote{This value is large enough to resolve the frequencies of the applied driving and at the same time it is short enough to avoid a pronounced time evolution of the nonstationary signal $x$.}.  As  the response $\hat R (f,t)$ converges more rapidly than fluctuations, the values of G are averaged only
on 20 quenches for each value of f.  
In figs.~\ref{fig:GGel}(a) and \ref{fig:GGel}(b) we plot the time evolution of $G'$ and $G''$ measured at $f=5$~Hz during the first 20 minutes after the quench. We can identify three different regimes:
\begin{itemize}
	\item[I.]{For 0 s $\le t \le$ 60 s, $G'$ is completely negligible whereas $G''$ increases continuously in time by a factor of almost two. In this aging regime, hightlighted in fig.~\ref{fig:GGel}(b), the gelatin droplet behaves as a purely viscous liquid even when the final temperature of the quench is below $T_{gel}$.}
	\item[II.]{For 60 s $\le t \le$ 200 s, $G''$ continuously increases and $G'$ starts to increase slowly, as plotted in fig.~\ref{fig:GGel}(b). This result shows that the droplet is in a transient regime towards the sol-gel transition where the gel network is not completely formed. Indeed, $G'$ is still much smaller and it increases slower than $G''$.}
	\item[III.]{For 200 s $\le t \le$ 1200 s, both $G'$ and $G''$ reach a logarithmic growth $\sim \log t$ where $G'$ increases faster than $G''$. This growth as $t$ increases is similar to that reported in macroscopic bulk measurements  \cite{normand,djabourov1,joly} but taking place much faster because of the smallness of the gelatin droplet. Then the system is actually undergoing gelation providing evidence that the percolating gel network is already formed.}
\end{itemize}

\begin{figure}
     \centering
     \subfigure[]{
          \includegraphics[width=.252\textwidth]{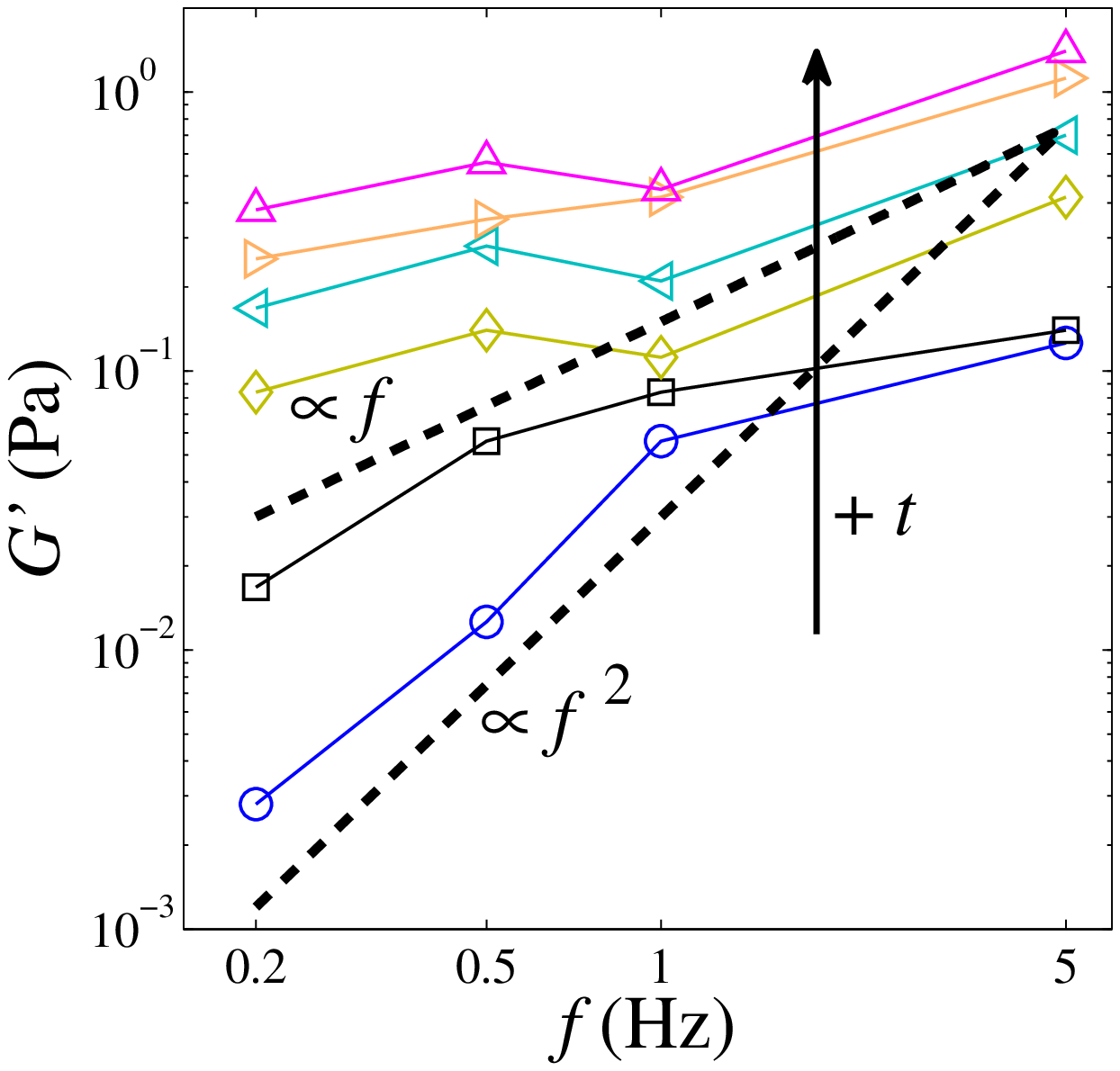}}
     \hspace{-0.275in}
     \subfigure[]{
          \includegraphics[width=.252\textwidth]{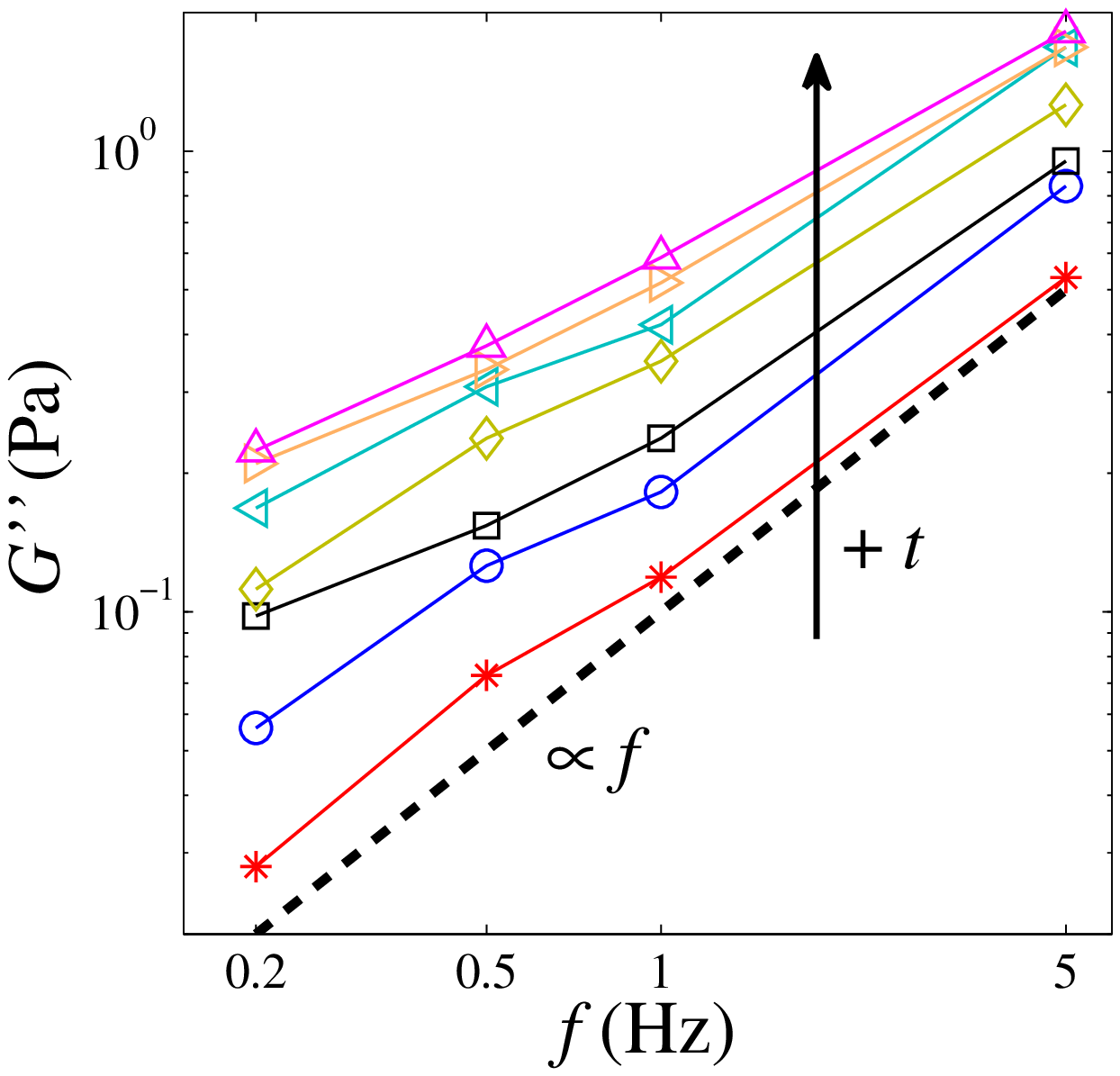}}\\
     \caption{(a) Storage and (b) loss modulus measured at different times after the quench: $t=60 \, (\ast),120 \,(\circ),180 \,(\Box), 300 \,(\Diamond), 600 \,(\triangleleft), 900 \,(\triangleright)$ and 1200~s~($\bigtriangleup$).}
     \label{fig:GGelf}
\end{figure}

In figs.~\ref{fig:GGelf}(a) and \ref{fig:GGelf}(b) we plot the low-frequency spectrum (0.2~Hz~$\le f \le $~5.0~Hz) of $G'$ and $G''$, respectively, at different times after the quench. In figs.~\ref{fig:GGelf}(a) the values of  $G'$ in regime I (0 s $\le t \le$ 60 s) are not reported as their mean values are close to zero and smaller than the statistical errors. Besides, in regime I   $G''$ is approximately proportional to $f$ [dashed line in fig.~\ref{fig:GGelf}(b)], and the dynamic viscosity of the droplet $G''/(2\pi f)$ is frequency-independent. Therefore, at this time scales we check that the fluid inside the droplet actually behaves as a Newtonian fluid with a vanishing relaxation time. In contrast, for $t$ larger than 60~s, $G'$ begins to grow. In fig.~\ref{fig:GGelf}(a) we plot as dashed lines the curves $\propto f$ and $\propto f^2$. As $f \rightarrow 0$, $G'$ displays an intermediate behavior between these two reference curves. Then, we conclude that $G'(f,t) \rightarrow 0$ as $f \rightarrow 0$ in regime II (60 s $\le t \le$ 200 s). On the other hand, in fig.~\ref{fig:GGelf}(b) we observe that $G''$ is roughly proportional to $f$. This behavior of $G$ can be approximately described by the Maxwell model for a viscoelastic fluid with a single relevant relaxation time $\tau_0$ and zero-shear viscosity $\eta_0$
\begin{equation}\label{eq:Maxwellmodel}
	G(f,t)=-\frac{2\pi \mathrm{i} f \eta_0(t)}{1+ 2\pi \mathrm{i} f \tau_0(t)}.
\end{equation}
{ 
Based on the Maxwell model, we estimate $\tau_0(t)$ and the zero-shear viscous drag coefficient $\gamma_0(t) = 6\pi r \eta_0(t)$ at aging time $t$ (see ref.\cite{gomez} for details)

As expected, $\tau_0$ is very small ($\approx 2$ ms)  during the first 60~s after the quench. 
Therefore, in regime I, the sol droplet can be regarded as a Newtonian fluid. Instead, for 60~s~$\le t \le 200$~s the mean value of $\tau_0$ becomes  four times the value found for the regime I. The values of $\tau_0$ must be compared  with the viscous relaxation time of the particle inside the optical trap: $\tau_k = {\gamma_0}/{k}$. 
Since $\tau_0/\tau_k \le 0.05$ even in regime II, the viscoelastic memory effects of the droplet taking place during $\tau_0$ do not significantly affect the Brownian motion of the particle during the first 200~s.}
Finally, for regime III (200 s $\le t \le$ 1200~s) $G''$ and mainly $G'$ exhibit a complex frequency dependence. Unlike the Maxwellian behavior ($G' \rightarrow 0$ as $f \rightarrow 0$), in this regime the value of $G'$ seems to remain non-zero as  $f \rightarrow 0$, as shown in fig.~\ref{fig:GGelf}(a). This result is in agreement with bulk measurements \cite{djabourov1,normand} that report a non-zero storage modulus of constant value at very low frequencies. This low-frequency behavior corresponds to that of an elastic solid with multiple relaxation time-scales.

\section{Heat flux}
\begin{figure}
     \centering
     \subfigure[]{
          \includegraphics[width=.36\textwidth]{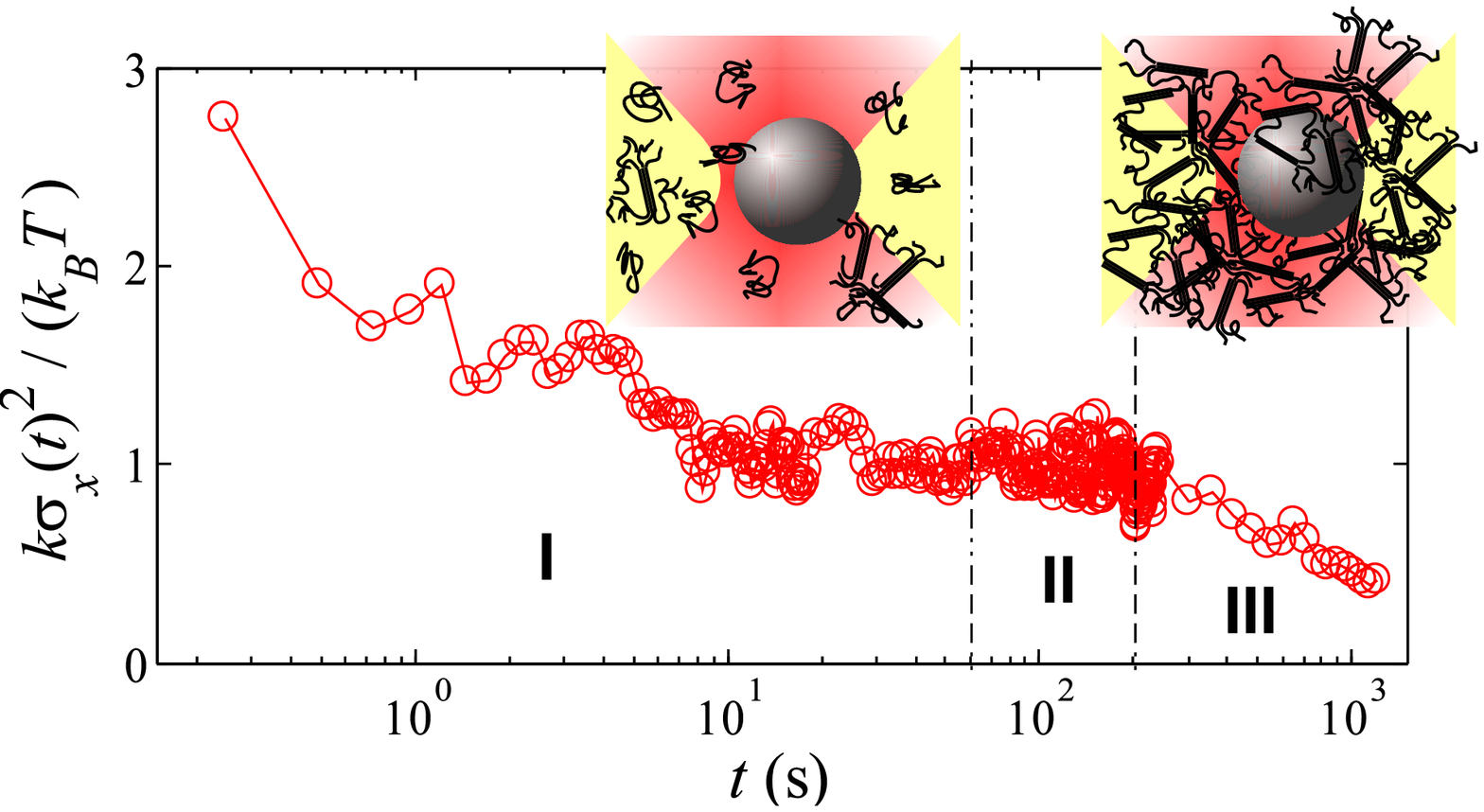}}\\
     \vspace{-0.175in}
     \subfigure[]{
          \includegraphics[width=.4\textwidth]{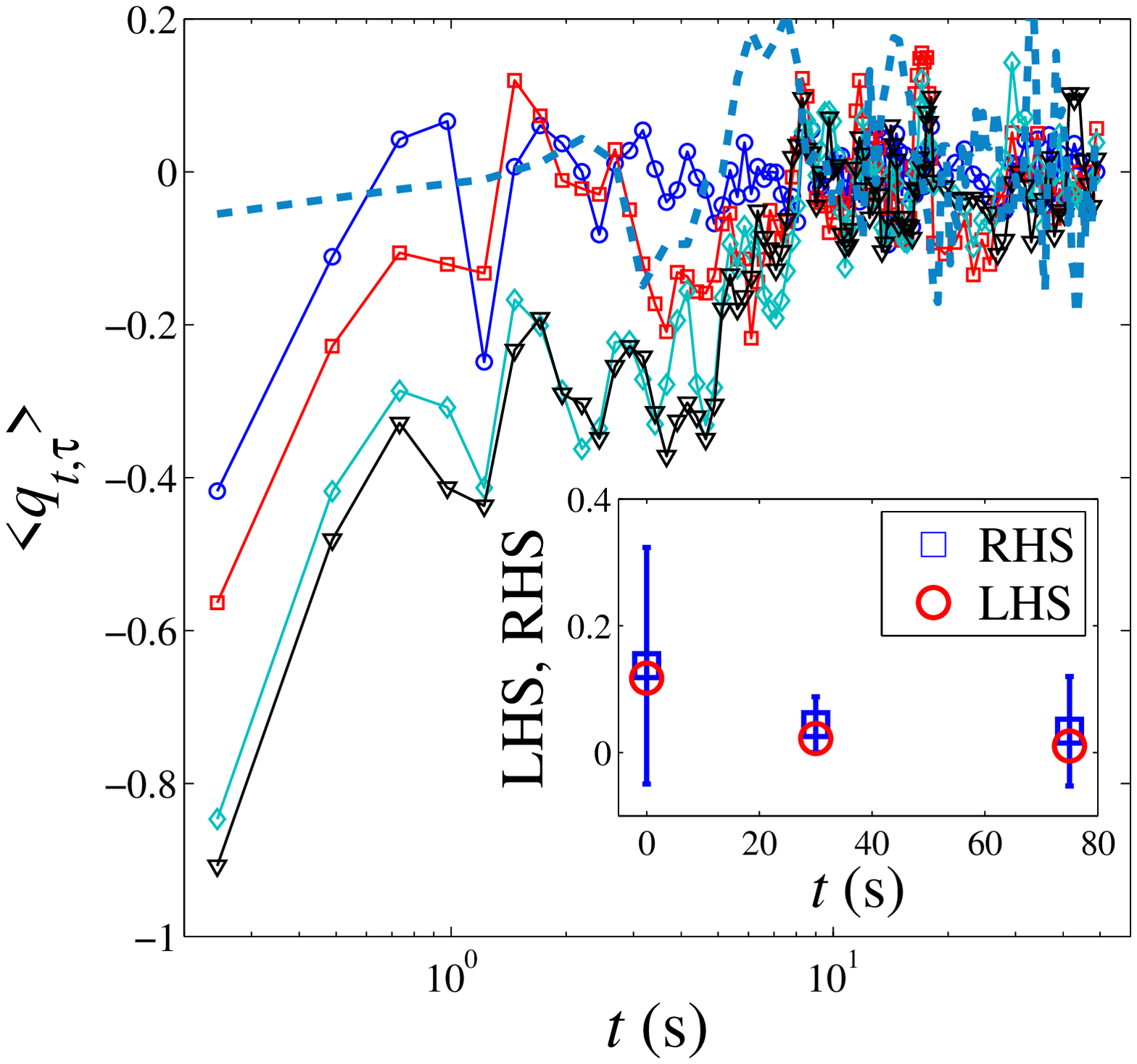}}\\
     \caption{{ (a) Time evolution of the variance of the particle position after the quench. The insets depict the possible configurations of the bead during the sol-gel transition of the surrounding gelatin. (b) Mean value of the normalized heat at different times $t$ after the quench in gelatin, for different values of the time lag: $\tau = 0.25$~s ($\circ$), 2.5~s ($\Box$), 10~s ($\Diamond$), 30~s ($\bigtriangledown$). The dashed line correspond to the normalized mean heat for the quench in glycerol at  $\tau = 30$~s. Inset: Comparison between the right-hand side (RHS) and left-hand side (LHS) of eq.~(\ref{eq:HaradaSasa}), normalized by $2 k_B T / k$, computed over a time window $[t,t+\Delta t]$ for $\Delta t=15$~s.}}
     \label{fig:meanheat}
\end{figure}

We now focus on the spontaneous energy fluctuations ($F=0$) of the Brownian particle inside the aging droplet  for the very first 200~s after the quench. Since the motion of the trapped particle is overdamped, its total energy at time $t$ is
\begin{equation}\label{eq:U}
	U(t) = \frac{1}{2}k x_t^2 + U_{stored}(t),
\end{equation}
where $U_{stored}(t)$ is the energy stored by the surrounding gelatin chains in the droplet until time~$t$. { By considering a non-Markovian overdamped Langevin equation for the dynamics of $x$ and eq.~(\ref{eq:Maxwellmodel}) as a model for the corresponding memory kernel \cite{jop}, the mean energy stored by the bath at  time $t$ can be estimated in this time interval (aging regimes I and II) by \cite{Solano_thesis}} 
\begin{equation}\label{eq:meanUstored}
	\langle U_{stored}(t) \rangle = \frac{\langle U(t) \rangle}{1+\frac{\tau_k(t)}{\tau_0(t)}}.
\end{equation}
{ 
As $\frac{\tau_0(t)}{\tau_k(t)}<0.05$ for the first $t<200$s, we deduce  from eq.~(\ref{eq:meanUstored}) that the energy stored by the bath constitutes in average less than 5\% of the total potential energy of the particle (see also ref.\cite{gomez})}. 
Hence, the instantaneous value of the total potential energy at time $t$ is given with an error smaller than 5\% by $U_t = k x_t^2/2$.
As there is no external force acting on the particle, 
the heat exchanged between the particle and the bath during the time interval $[t,t+\tau]$ is equal to the variation 
$\Delta U_{t,\tau} = U_{t+\tau} - U_t$  of the total energy of the particle\footnote{This is not valid during the aging regime III because in such a case we have to take into account the particle history between $t$ and $t+\tau$ determined by the exact form of the shear modulus $G(f,t)$.} (see ref. \cite{sekimoto}). Specifically:
{ $Q_{t,\tau} = \Delta U_{t,\tau} ={k}(x_{t+\tau}^2-x_t^2)/2$,}
where a positive (negative) value of $Q_{t,\tau}$ represents a heat fluctuation from (to) the bath to (from) the bead. 
Then, the mean heat transferred during $[t,t+\tau]$ can be written as
\begin{equation}\label{eq:meanheat}
	\langle Q_{t,\tau} \rangle = \frac{k}{2}[\sigma_x({t+\tau})^2-\sigma_x(t)^2] ,
\end{equation}
where $\sigma_x(t)^2$ is the ensemble variance of the spontaneous fluctuations of $x$ at time $t$. In ref.~\cite{gomez} we show that the fluctuations of $x$ are Gaussian and $\sigma_x^2(t)$ is a monotonically decreasing function of $t$. This time evolution is plotted in fig.~\ref{fig:meanheat}(a) for the three different aging regimes identified by active microrheology. During the first $\approx 20$~s, $\sigma_x(t)^2 > k_B T / k$, which can be interpreted as the result of both thermal and non-thermal stochastic forces on the particle due to the transient formation of the gel network, as schematized in the inset of fig.~\ref{fig:meanheat}(a). The dimensionless heat $q_{t,\tau}=Q_{t,\tau}/(k_B T)$, computed using eq.~(\ref{eq:meanheat}),  is plotted in fig.~\ref{fig:meanheat}(b).  Because of the relaxation
of $\sigma_x(t)^2$, we observe the existence of a mean heat flux from the particle to the
surroundings on the time scale $\tau$. The absolute value of the mean heat increases as the measurement time $\tau$ increases and the maximum value $|\langle q_{t,\tau} \rangle |\approx 1$ takes place at $t=0$ s.
Note that at thermal equilibrium the mean heat would be $\langle q_{t,\tau} \rangle = 0$ for all $t$ and $\tau$. 
Non-negligible values of the mean heat persist for several seconds after the quench.
Nevertheless, as $t$ increases, $|\langle q_{t,\tau} \rangle |$
decreases becoming experimentally undetectable for
$t \gtrsim$ 20 s. The non-vanishing mean heat flux $\langle q_{t,\tau} \rangle$ is a signature of the broken detailed balance of the particle dynamics due to the assembling gelatin chains. However, as the bath is undergoing aging the time scales of the dynamics slow down. Then the rate at which the heat flows from the system to the environement becomes undetectable by the Brownian probe, which results in an apparent equilibrium-like behavior for $t \gtrsim 20$~s.  In order to check for possible experimental artifacts, we repeated the same quench experiment using a Newtonian fluid (glycerol 60~wt~\% in water). At this concentration the glycerol solution has the same viscosity of the initial
sol phase of gelatin ($9 \times 10^{-3}$~Pa~s) but with no sol-gel transition in the same temperature range. For this fluid, the measured mean heat flux is plotted for comparison in fig.~\ref{fig:meanheat}. In this case we verify that the particle quickly equilibrates with the bath after the quench so that $\langle q_{t,\tau}\rangle = 0$ within the experimental accuracy.


In ref.~\cite{gomez} we show that the heat fluctuations verify the FT. In analogy with a system kept in a { non-equilibrium \emph{steady} state} in contact with two reservoirs at unequal temperatures \cite{Jarzynski_2004,lecomte,visco,bodineau,piscitelli,saito,kundu,evans}, the quantity 
\begin{equation}\label{eq:entropyGel}
	\Delta S_{t,\tau}= \frac{k_B}{k}\left[\frac{1}{\sigma_x(t)^2}-\frac{1}{\sigma_x(t+\tau)^2}\right] Q_{t,\tau},
\end{equation}
represents the total entropy produced by the breakdown of
the time-reversal symmetry due to the nonstationarity of the bath after the quench. { By taking the ensemble average of eq.~(\ref{eq:entropyGel}) and using eq.~(\ref{eq:meanheat}), one finds that the mean heat $\langle Q_{t,\tau}\rangle$ actually quantifies the mean total entropy production between time $t$ and $t+\tau$: $\langle  \Delta S_{t,\tau}\rangle  \propto { \langle Q_{t,\tau}\rangle^2 \over \sigma_x(t+\tau)^2 \sigma_x(t)^2}$.}

\begin{figure}
     \centering
     \subfigure[]{
          \includegraphics[width=.25\textwidth]{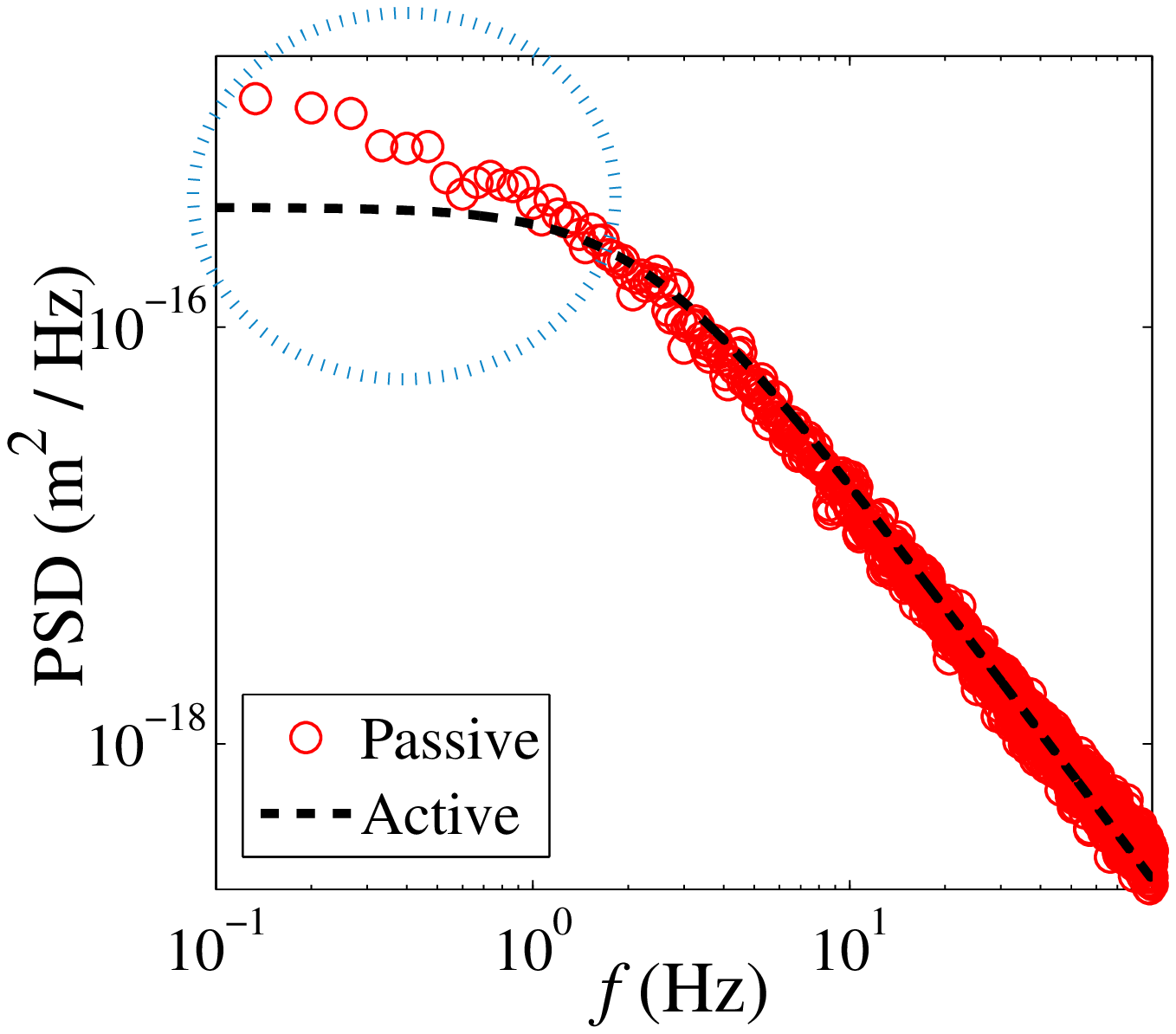}}
     \hspace{-0.27in}
     \subfigure[]{
          \includegraphics[width=.25\textwidth]{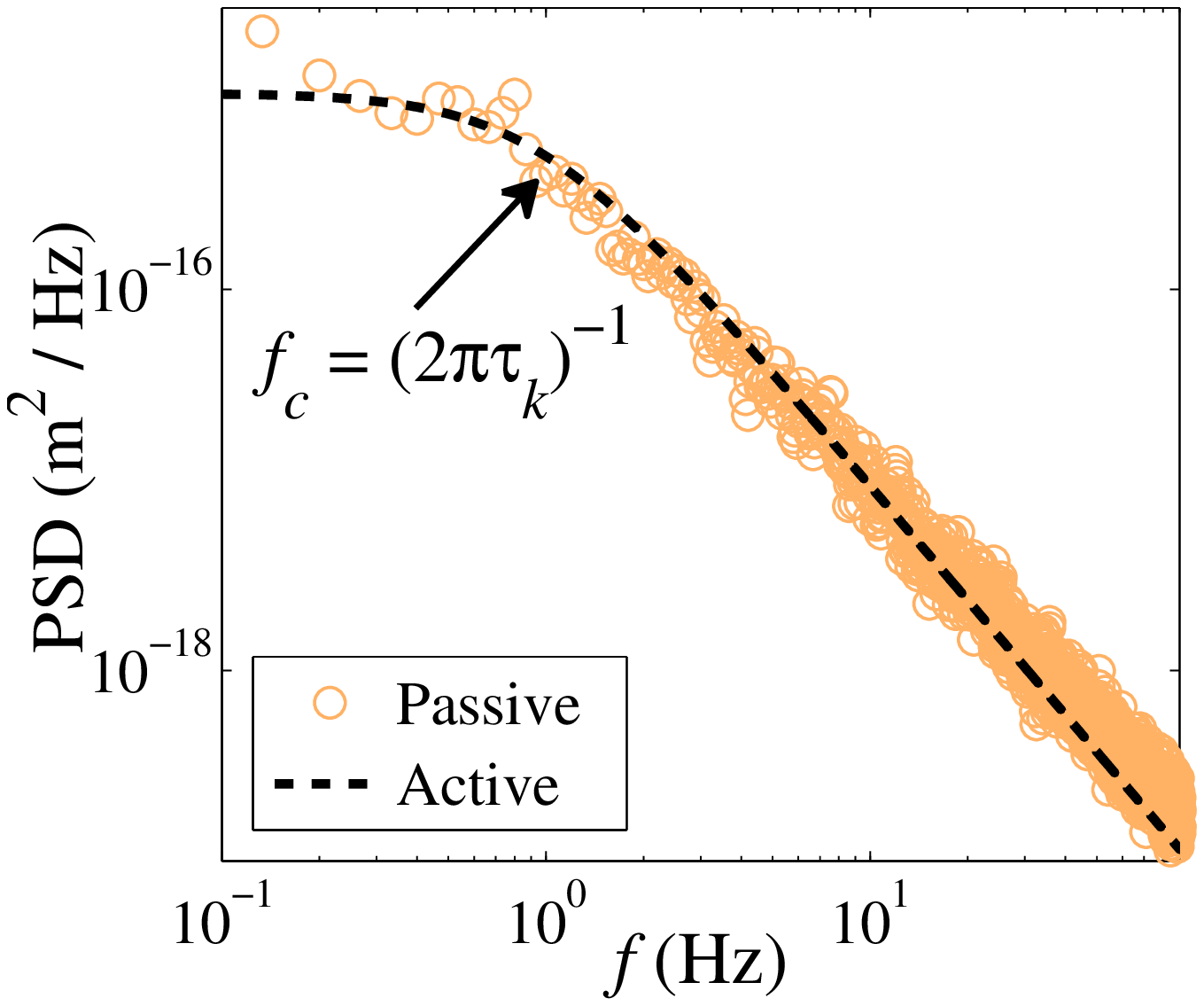}}\\
\vspace{-0.2in}
\subfigure[]{
          \includegraphics[width=.25\textwidth]{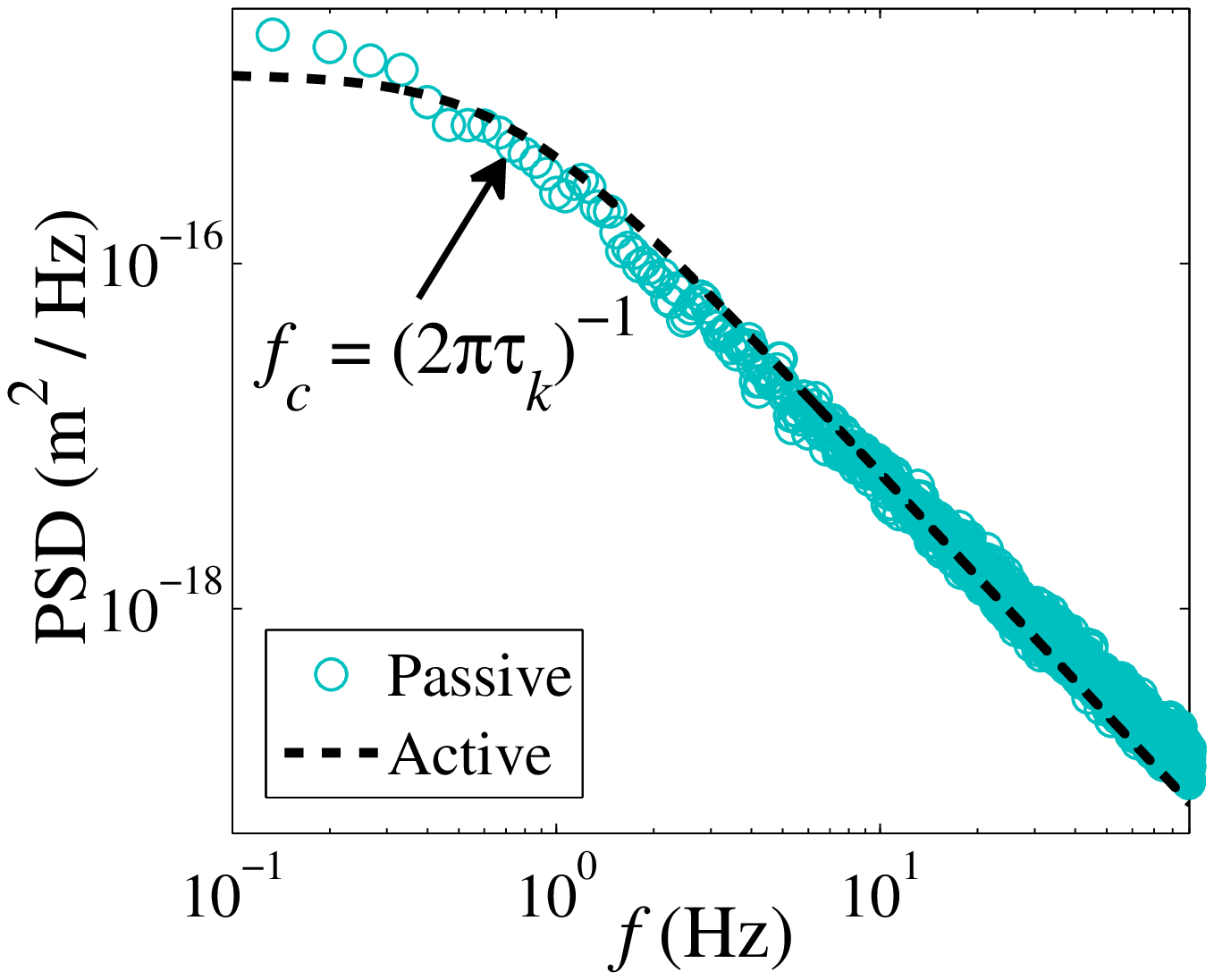}}
     \hspace{-0.27in}
     \subfigure[]{
          \includegraphics[width=.24\textwidth]{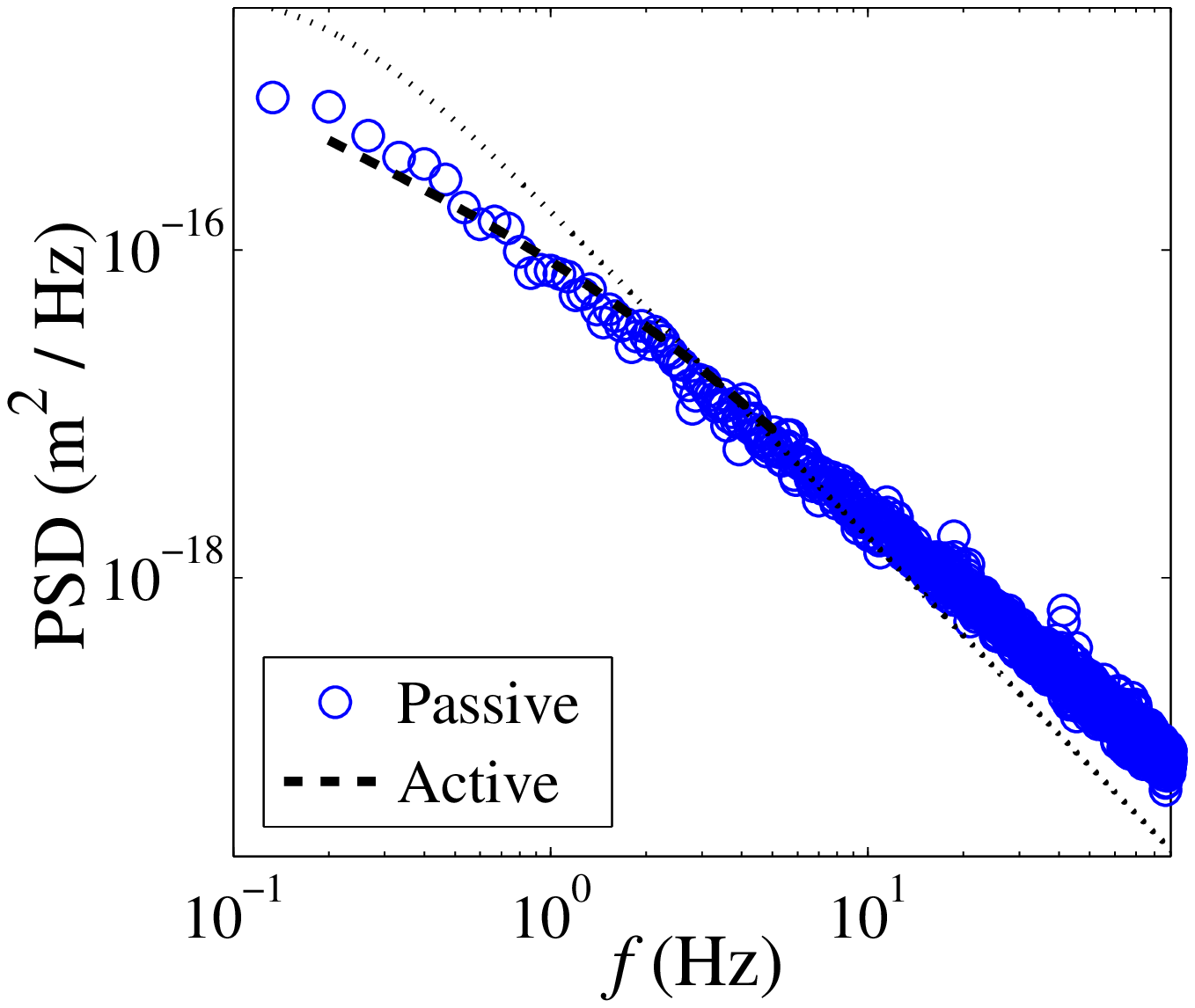}}\\
     \caption{{ Passive and active spectra {\color{red} computed on the interval [$t,t+\Delta t$] at different times $t$ } after the quench: (a) [0~s, 15~s], the dotted circle points out the low-frequency deviation from eq.~(\ref{eq:FDTGel}); (b) [30~s, 45~s], the arrow indicates the position of the frequency $f_c$; (c) [75~s, 90s]; and (d) [1200~s, 1215~s]. The dotted line corresponds the Lorentzian curve obtained  using eq.~(\ref{eq:ImRGelI}) ($G'=0$).}} 
     \label{fig:spectraGel}
\end{figure}

\section{Generalized fluctuation-dissipation relation}
We now discuss the relation between entropy production, spontaneous fluctuations and linear response of the trapped particle in this out-of-equilibrium aging bath. It should be expected that there is a strong connection between the deviations from the equilibrium FDT and the extent of the broken detailed balance, quantified by $\langle Q_{t,\tau}\rangle$. If the entropy production takes place at a very slow rate, then the particle in the aging bath should exhibit an equilibrium-like behavior, see the experiments on FDT in colloidal glasses of refs.~\cite{jabbari,jop}. On the other hand, if the rate is fast enough, then currents are non-negligible and such deviations should be significant like in the non-equilibrium steady state experiments of refs.~\cite{gomez1,mehl}. In the present case of gelatin, we found that $|\langle Q_{t,\tau} \rangle|$ is comparable to $k_B T$ during the first 20~s after the quench. The rate of this heat flux slows down as the gelatin ages. 
Therefore, a significant deviation of the equilibrium FDT is expected to occur for the position of the trapped particle during the aging regime I of the gelatin droplet. In contrast, in regimes II and III the FDT must be apparently satisfied with the temperature $T$ of the thermal  bath. Indeed, we show in the following that our experimental results display such a kind of behavior.

As we performed active microrheology at a fixed excitation frequency $f$, it is more convenient to study the fluctuations and linear response in the frequency domain. The Fourier transform $\hat{R}$ of the linear response function of $x$ to the perturbative time-dependent force $F$ is related to the shear modulus $G$ of the gelatin droplet by
\begin{equation}\label{eq:RG}
	\hat{R}(f,t) = \frac{1}{6 \pi r G^*(f,t)+k},
\end{equation}
where $G^*$ is the complex conjugate of $G$. 
Then, the imaginary part of eq.~(\ref{eq:RG}) is 
\begin{equation}\label{eq:ImRGel}
	\mathrm{Im}\{ \hat{R}(f,t) \} = \frac{6\pi r G''(f,t)}{[k+6\pi r G'(f,t)]^2+[6\pi r G''(f,t)]^2}.
\end{equation}
At thermal equilibrium, the quantity {  given in eq.~(\ref{eq:ImRGel})} {\color{red} does not depend on time} and  is related to the power spectral density $\langle |\hat{x}(f,t)|^2 \rangle$ of the spontaneous fluctuations of $x$ in the absence of an external force ($F=0$). This relation is provided by the FDT and reads
\begin{equation}\label{eq:FDTGel}
	\langle |\hat{x}(f,t)|^2 \rangle = \frac{2k_B T}{\pi f} \mathrm{Im}\{ \hat{R}(f,t) \}.
\end{equation}
 {\color{red}
During the non-equilibrium gelation of the gelatin droplet, eq.\ref{eq:FDTGel} does not necessarily hold because as we have already seen $\hat R$ is a function of both $t$ and $f$. Thus  we study the relation between { the \emph{passive} spectrum} $\langle |\hat{x}(f,t)|^2 \rangle$ and { the  \emph{active} spectrum} $2 k_B T \mathrm{Im}\{ \hat{R}(f,t) \} /(\pi f)$,  as a function of $t$}. { For this purpose we compute independently these quantities over a time interval $[t, t+\Delta t]$ of length $\Delta t = 15$~s at time $t$ after the quench in the different aging regimes.}

In fig.~\ref{fig:spectraGel}(a) we plot $\langle |\hat{x}(f,t)|^2 \rangle$ for $t = 0$~s, where a strong mean heat flux takes place { (regime I)}. On the other hand, taking into account the {  negligible values of $\tau_0$ in this regime, $\mathrm{Im}\{ \hat{R}\}$ can be written as
\begin{equation}\label{eq:ImRGelI}
	\mathrm{Im}\{\hat{R}(f,t)\} = \frac{1}{2\pi \gamma_0(t)} \frac{f}{f^2+f_c(t)^2},
\end{equation}
where $f_c(t)$ is related to the viscous relaxation time $\tau_k(t)=\gamma_0(t)/k$ of the trapped particle at time $t$ by: 
\begin{equation}
	f_c(t) = \frac{1}{2\pi \tau_k(t)}.
\end{equation}
In fig.~\ref{fig:spectraGel}(a) we also plot the active spectrum $2 k_B T \mathrm{Im}\{ \hat{R} \}/(\pi f)$ computed using eq.~(\ref{eq:ImRGel}). This spectrum  has a Lorentzian profile. As expected, an important deviation of eq.~(\ref{eq:FDTGel}) is observed for frequencies $f \lesssim 1$~Hz. Note that for frequencies $f \gtrsim 1$~Hz eq.~(\ref{eq:FDTGel}) is verified, indicating that the deviation is actually due to the slow structural assemblage of the gel\footnote{{ Similar low-frequency violations of the FDT are found in non-equilibrium active systems, \emph{e.g.} biological polymer networks \cite{mizuno}.}}. The spectral curves of fig.~\ref{fig:spectraGel}(a) provide information on the time scales at which this phenomenon perturbs the Brownian probe: $\gtrsim 1$~s. The extent of the deviation, i.e. the difference between the area below $\langle |\hat{x}(f,t)|^2 \rangle $ minus the area below $2 k_B T \mathrm{Im}\{ \hat{R} \}/(\pi f)$ in fig.~\ref{fig:spectraGel}(a), is
\begin{equation}\label{eq:HaradaSasa}
	\int_{0}^{\infty} \left[ \langle |\hat{x}(f,t)|^2 \rangle  - \frac{2k_B T}{\pi f} \mathrm{Im}\{ \hat{R}(f,t) \} \right]\,\mathrm{d}f =  \frac{2   \overline{ Q}_{t,\Delta t} }{k},
\end{equation}
{ 
where  $\overline{ Q}_{t,\Delta t} $ can be interpreted  as {\color{red}  the heat that the system dissipates in average during the time interval  $ [t, t+\Delta t] $}
\cite{Solano_thesis}.  This  interpretation is justified because, using  eqs.~(\ref{eq:HaradaSasa}), (\ref{eq:ImRGelI}) and (\ref{eq:meanheat}),  it can be easily shown \cite{Solano_thesis} that indeed $\overline{ Q}_{t,\Delta t}  = -{1\over \Delta t}\int_{t}^{t+ \Delta t} \, \langle Q_{t',\bar\tau} \rangle  \, \mathrm{d}t'  $ with  $\bar \tau> 30$~s, {\color{red} such that at time $t'+\bar \tau$ the systems has reached equilibrium for all times after the quench} [see fig.\ref{fig:meanheat}(a)].  We used the experimental data to compute independently the left-hand side (LHS) and the right-hand side (RHS)
of eq~(\ref{eq:HaradaSasa}) and we find that they are equal within error bars,} { as shown in the inset of fig.~\ref{fig:meanheat}(b)}. }

Thus similarly to the GFDR for non-equilibrium steady states derived in \cite{harada}, eq.~(\ref{eq:HaradaSasa}) hightlights the role of heat currents (or equivalently the entropy production). In the present case, the right-hand side of eq.~(\ref{eq:HaradaSasa}) quantifies the heat excess that must irreversibly flow from the particle to the bath in order for the system to reach equilibrium at temperature $T$. Thus, the \emph{violation} of the FDT (\ref{eq:FDTGel}) can be interpreted as a measure of the heat that must be dissipated to reach an equilibrium state. This is in close analogy to the power dissipated into the medium to keep a non-equilibrium steady state in the GFDR of ref. \cite{harada}. In fig.~\ref{fig:spectraGel}(b) we plot the passive and active spectra in the regime I for the $t = 30$~s. We check that in this case the equilibrium-like eq.~(\ref{eq:FDTGel}) is satisfied with the final equilibrium temperature $T=27.2^{\circ}$C. This results is in good agreement with the fact that the mean heat flux becomes undetectable by the Brownian particle for $t>20$~s. 

Fig.~\ref{fig:spectraGel}(c) displays the passive power spectrum of $x$ for $t = 75$~s, (regime II) where $G'$ starts to increase. On the other hand, taking into account that the fluid can be characterized by a single relaxation time $\tau_0 \le 0.05 \tau_k$, the active term can be approximated with an error smaller than 5\% by eq.~(\ref{eq:ImRGelI}).
Using this expression with the experimental values of $\gamma_0$ and $f_c$, in fig.~\ref{fig:spectraGel}(c) we plot the active term $2 k_B T \mathrm{Im}\{ \hat{R} \}/(\pi f)$. In this case, the agreement between the left- and the right-hand side of eq.~(\ref{eq:FDTGel}) is much better than in fig.~\ref{fig:spectraGel}(a).  Hence, we verify that the system is  relaxing  to an equilibrium-like behavior. 

Finally, in fig.~\ref{fig:spectraGel}(d) we plot the passive and active spectra for $t =$~1200~s. In this case the large viscoelasticity of the gelatin droplet gives rise to the highly non-Lorentzian shape of both terms.
For comparison, we also plot the Lorentzian profile (dotted line) that would be obtained using equation (\ref{eq:ImRGelI}) without taking into account the contribution of the storage modulus $G'$.
In the absence of an accurate model for $G$ in regime III, we only plot the active term in the frequency range 0.2~Hz$\le f \le5$~Hz at which the sinuosoidal external force was applied to the particle. We verify that an equilibrium-like fluctuation-dissipation relation (\ref{eq:FDTGel}) holds for the particle position $x$ because of the smallness of the heat fluxes as the gelatin droplet ages.

\section{Conclusion}
We have experimentally measured the fluctuations of
the position of a trapped Brownian particle in a
non-stationary bath, \emph{i.e.} an aging gel after a very fast
quench. This simple experiment has allowed us to understand several concepts formulated originally for
non-equilibrium steady states that can be naturally extended to non-stationary systems slowly relaxing towards an equilibrium state. In particular, we provide evidence that the extent of the \emph{violation} of the equilibrium FDT is directly related to the entropy that must be produced in order for the system to reach equilibrium. Therefore, as the system ages, the relation between fluctuations and linear response behaves like in equilibrium. This approach to fluctuation-dissipation, alternative to the concept of effective temperature \cite{cugliandolo}, can be exploited to study small heat fluxes in experiments on more complex aging systems.

\acknowledgments


\begin{thebibliography}{0}
 
 
 \bibitem{gallavotti}
\Name{Evans D. J. \emph{et al.}} \REVIEW{Phys. Rev. Lett.} {71} {2401} {1993}. 
\Name{Gallavotti G. \and  Cohen E. G. D.} \REVIEW{J. Stat. Phys.} {80} {931} {1995}.

\bibitem{cugliandolo} 
\Name{Cugliandolo L. F., Kurchan J., \and Peliti L.} 
\REVIEW{Phys. Rev. E}{55}{3898} {1997}.


\bibitem{lippiello} 
\Name{E. Lippiello, F. Corberi, \and M. Zannetti}
\REVIEW{Phys. Rev. E}{71}{036104}{2005}.

\bibitem{baiesi0} 
\Name{M. Baiesi, C. Maes, \and B. Wynants} 
\REVIEW{Phys. Rev. Lett.}{103}{010602}{2009}.

\bibitem{Seifert_2010} 
\Name{U. Seifert \and T. Speck}
\REVIEW{EPL}{89}{10007}{2010}.

\bibitem{verley} 
\Name{Verley G., Chetrite R.  \and Lacoste D.}
\REVIEW{J. Stat. Mech.}{}{2011}{P10025}.

\bibitem{ciliberto} 
\Name{Ciliberto S., Joubaud S. \and Petrosyan A.} 
\REVIEW{J. Stat. Mech.}{}{2010}{P12003}.

\bibitem{berthier} 
\Name{Berthier L. \and Biroli G.} 
 \REVIEW{Rev. Mod. Phys.} {83}{587} {2011}.

\bibitem{crisanti} 
\Name{Crisanti A. \and Ritort F.}
\REVIEW{EPL}{66}{253}{2004}.

\bibitem{ritort} 
\Name{Ritort F.}
\REVIEW{J. Phys. Chem. B}{108}{6893}{2004}.

\bibitem{zamponi}
  \Name{Zamponi F. \emph{et al.}}
  \REVIEW{J. Stat. Mech.}{}{2005}{P09013}.

\bibitem{chetrite} 
\Name{R. Chetrite}
\REVIEW{Phys. Rev. E}{80}{051107}{2009}.

\bibitem{djabourov1} 
\Name{Djabourov M, Leblond J., \and Papon P.}
\REVIEW{J. Phys. France}{49}{319}{1988}.
\Name{Djabourov M, Leblond J., \and Papon P.}
\REVIEW{J. Phys. France}{49}{333}{1988}.




\bibitem{normand}
\Name{Normand V. \emph{et al.}}
\REVIEW{Macromolecules}{33}{1063}{2000}.


\bibitem{gomez} 
\Name{Gomez-Solano J. R., Petrosyan A., \and Ciliberto S.}
\REVIEW{Phys. Rev. Lett.}{106}{200602}{2011}.

\bibitem{peterman}
\Name{Peterman E. J.  G. \emph{et al.}} 
\REVIEW{Biophys. J} {84} {1308}  {2003}.


\bibitem{Solano_thesis} 
More details can be found in
\Name{R.Gomez-Solano}
\REVIEW{PhD Thesis-ENSL , http://tel.archives-ouvertes.fr }{}{2011}{}  and in a longer report
R.  Gomez-Solano et all in preparation.  


\bibitem{resolution} 
\Name{Andrieux D. \emph{et al.}}
\REVIEW{J. Stat. Mech.}{}{2008}{P01002}.

\bibitem{jop}
  \Name{Jop P., Gomez-Solano J. R., Petrosyan A. \and Ciliberto S.}
  \REVIEW{J. Stat. Mech.}{}{2009}{P04012}.

\bibitem{joly} 
\Name{Joly-Duhamel C. \emph{et al.}}
\REVIEW{Langmuir} {18}{7158} {2002}.

\bibitem{sekimoto} 
\Name{Sekimoto K.} \REVIEW{Prog. Theor. Phys. Suppl.} {130} {17} {1998}. 
\Name{Van Zon R. \and Cohen E. G. D.} \REVIEW{Phys. Rev. Lett.} {91} {110601} {2003}.

\bibitem{Jarzynski_2004} 
\Name{Jarzynski C. \and Wojcik D. K.} 
\REVIEW{Phys. Rev. Lett} {92} {230602}{2004}.

\bibitem{lecomte} 
\Name{Lecomte V., Racz Z., \and van Wijland F.}
\REVIEW{J. Stat. Mech.} {}{2005}{P02008}.

\bibitem{visco} 
\Name{Visco P.} 
\REVIEW{J. Stat. Mech.}{}{2006}{P06006}.

\bibitem{bodineau} 
\Name{Bodineau T. \and Derrida B.}
\REVIEW{C. R. Physique}{8}{540}{42007}.

\bibitem{piscitelli} 
\Name{Piscitelli A., Corberi F., \and Gonnella G.}
\REVIEW{J. Phys. A: Math. Theor.}{41} {332003}{2008}.

\bibitem{saito} 
\Name{Saito K. \and Dhar A.}
\REVIEW{Phys. Rev. E} {83}{041121} {2011}.

\bibitem{kundu} 
\Name{Kundu A., Sabhapandit S., \and Dhar A.} 
\REVIEW{J. Stat. Mech.}{}{2011} {P03007}.


\bibitem{evans}
\Name{Evans D. , Searles D. J.  \and Williams S. R.} \REVIEW{J. Chem. Phys.} {132} {024501}{2010}.

\bibitem{jabbari} 
\Name{Jabbari-Farouji S. \emph{et al.}}
\REVIEW{EPL}{84}{20006}{2008}.


\bibitem{gomez1}
  \Name{Gomez-Solano J. R. \emph{et al.}}
  \REVIEW{Phys. Rev. Lett.}{103}{040601}{2009}.

\bibitem{mehl}
  \Name{Mehl. J., Blickle V., Seifert U., \and Bechinger C.}
  \REVIEW{Phys. Rev. E}{82}{ 032401}{2010}.

\bibitem{mizuno}
\Name{Mizuno D. \emph{et al.}}
\REVIEW{Science}{315}{370}{2007}.

\bibitem{harada} 
\Name{Harada T. \and Sasa S.-I.}
\REVIEW{Phys. Rev. Lett.}{95}{130602}{2005}.





\end{thebibliography}
\end{document}